# Visualization of Unconventional Rashba Band and Vortex Zero Mode in Topological Superconductor Candidate AuSn$_4$

Yuhan Ye[1,2,#], Rui Song[3,#], Hongqin Xiao[1,2,#], Guoyu Xian[1,2], Hui Guo[1,2,4], Haitao Yang[1,2,4], Hui Chen[1,2,4,*] and Hong-Jun Gao[1,2,4]

[1] *Beijing National Center for Condensed Matter Physics and Institute of Physics, Chinese Academy of Sciences, Beijing 100190, PR China*

[2] *School of Physical Sciences, University of Chinese Academy of Sciences, Beijing 100190, PR China*

[3] *Science and Technology on Surface Physics and Chemistry Laboratory, Mianyang 621908, PR China*

[4] *Hefei National Laboratory, 230088 Hefei, Anhui, PR China*

[#] These authors contributed equally to this work

[*] Correspondence to: hchenn04@iphy.ac.cn

# ABSTRACT

Topological superconductivity (TSC) is a promising platform to host Majorana zero mode (MZM) for topological quantum computing. Recently, the noble metal alloy $AuSn_4$ has been identified as an intrinsic surface TSC. However, the atomic visualization of its nontrivial surface states and MZM remains elusive. Here, we report the direct observation of unconventional surface states and vortex zero mode at the gold (Au) terminated surfaces of $AuSn_4$, by ultra-low scanning tunneling microscope/spectroscopy. Distinct from the trivial metallic bulk states at tin (Sn) surfaces, the Au terminated surface exhibits pronounced surface states near Fermi level. Our density functional theory calculations indicate that these states arise from unconventional Rashba bands, where two Fermi circles from different bands share identical helical spin textures, chiralities, and group velocities in the same direction. Furthermore, we find that although the superconducting gap, critical temperature, anisotropic in-plane critical field are almost identical on Au and Sn terminated surfaces, the in-gap bound states inside Abrikosov vortex cores show significant differences. The vortex on Sn terminated surfaces exhibits a conventional Caroli-de Gennes-Matricon bound state while the Au surface shows a sharp zero-energy core state with a long non-splitting distance, resembling an MZM in a non-quantum-limit condition. This distinction may result from the dominant contribution of unconventional Rashba bands near Fermi energy from Au terminated surface. Our results provide a new platform for studying unconventional Rashba band and MZM in superconductors.

The pursuit of topological superconductors represents an essential pathway towards potential of fault-tolerant quantum computing[1, 2]. While the roster of materials boasting nontrivial topological bands encircling the Fermi level is rapidly expanding[3], only a select few harbor superconducting properties. TSs can manifest through the discovery of *p*-wave superconductors or by inducing *s*-wave superconductivity on the surface of quantum materials endowed with symmetry-protected topological surface states. The latter has been proposed in several hybrid structures such as Pb-$Bi_2Te_3$ [4] and $Bi_2Se_3$ thin film on $NbSe_2$ substrate[5]. These materials exhibit a full pairing gap in the bulk, accompanied by gapless surface states and tied to the appearance of zero-energy excitations, known as Majorana zero mode (MZM) [6-9]. In addition to hybrid structure, the signature of MZMs have been also observed in a few single superconducting compounds with topological surface states such as $Cu_xBi_2Se_3$[10, 11], kagome superconductor $CsV_3Sb_5$[12, 13], transition metal dichalcogenide 2M-$WS_2$ [14, 15], iron-based superconductors including $FeTe_xSe_{1-x}$[16-21], $CaKFe_4As_4$[22], $LiFeAs$[23, 24] and $Li_{0.84}Fe_{0.16}OHFeSe$[25]. To date, exploring new superconductors hosting the MZMs, stands as a formidable challenge in the realm of condensed matter physics.

In addition to the topological surface states, combination with unconventional Rashba band and superconductivity is another promising avenue to realize MZM. The Fermi surface of Rashba band is similar to that of topological surface states: the former is consisted of two closed circles while the latter has only one. For conventional two-band Rashba system, the opposite spin textures of the inner and outer Fermi circles result in opposite contribution to the spin-to-charge conversion[26, 27]. This partial compensation effect gets rid of the possibility for the system to hold topological superconductivity[28]. However, strong Rashba spin-orbit coupling between two sets of two-band Rashba system provide two bands with same spin textures[29-31]. In this scenario, two Fermi circles from different bands share the same helical spin textures with the same chiralities and group velocities with the identical direction. Such unconventional Rashba band structure significantly enhances the spin-to-charge conversion, resembling the condition of topological surface states. Recently, $AuSn_4$, a type of noble metal alloys $AB_4$ [32-42] (*A* = Au, Pt, and Pd, *B* = Sn and Pb), has been claimed to exhibit two-dimensional superconductivity[38-40], surface *p*-wave superconductivity[40] and unconventional Rashba split[40]. Nevertheless, the atomic scale study of topological surface states and MZMs remains elusive.

In this work, we report the direct observation of nontrivial surface states and robust zero-energy vortex bound states on the Au-terminated surface of $AuSn_4$ by utilizing ultra-low temperature scanning tunneling microscopy/spectroscopy (STM/S) equipped with external magnetic field combined with first-principles calculations. We distinguish between Au and Sn terminated surfaces and find that unconventional Rashba bands are dominant at Au surfaces, whereas the Sn surface exhibits metallic bulk states. In the superconducting state, the superconducting gap, critical temperature and magnetic field are almost the same for the two surfaces. Remarkably, by applying external magnetic field perpendicular to the surface, we observe distinct vortex bound states inside the Abrikosov vortices of An and Sn surfaces. The Sn surface shows a broadened zero-energy core state with a short decay length, while the vortex on Au surface exhibits a sharp zero-bias bound states with long decay length, resembling the MZM in non-quantum-limit condition. The MZM signature may result from the dominant contribution of unconventional Rashba bands near Fermi surface on the Au surface.

The AuSn$_4$ crystal, which has an orthorhombic crystal structure (space group *Aba2*, *a* = *b* = 6.476 Å, *c* = 11.666 Å, crystal growth details see Experimental Method and Figure S1), consisting of periodic stacking tri-layers: Sn-Au-Sn (Figure 1(a)). For each stacking tri-layer, the Au layer is sandwiched by two Sn layers. Therefore, there are two possible terminated surfaces, i.e. Au and Sn surfaces.

In the STM topographic images, we observe two types of terminated surfaces on the as-cleaved AuSn$_4$ crystal (Figure 1(b)). There are several distinct topographic features for the two surfaces: (1) Type-I surfaces show stripe-like patterns with randomly distributed adatoms, while type-II surfaces show ordered square lattice with randomly distributed vacancies (Figure 1(c) and Figure S2). (2) The Type-I surfaces are observed much more frequently (98%) compared to type-II surfaces (2%), indicating that type-I surfaces are the most favorable cleave plane. (3) The smallest step height between two adjacent Type-I terraces is ~0.62 nm, consistent with the distance between two equivalent layers in two adjacent tri-layer units. In contrast, the height between type-I and type-II terraces is about 0.16 nm. According to the atomic model, the type-II terminated surfaces are attributed to the Au layer sandwiched by two Sn layers (lower panel in Figure 1(b)). The stripe patterns at type-I surface (Figure S3) are previously associated with unidirectional charge density waves (CDW) on the Sn terminated surfaces[40], further supporting the termination identification. Additionally, we find the unidirectional CDW on the Sn surface is not uniformly oriented. The direction and corresponding electronic states of unidirectional CDW pattern occasionally vary.

In addition to their different topographic features, the Au and Sn terminated surfaces exhibit dramatically distinct electronic states, as evidenced by the spatially-averaged d*I*/d*V* spectra obtained from both surfaces (Figure 1(d)). The Sn surface displays metallic behavior with a V-shape feature centered at Fermi level ($E_F$), indicative of contributions primarily from bulk electronic states. In contrast, the Au surface show several pronounced peaks near $E_F$ at energies of $E \approx$ -352 meV ($P_0$), -109 meV ($P_1$), 24 meV ($P_2$) and 235 meV ($P_3$), respectively. These peaks shift to lower energy across the adatoms or defects, suggesting their origin from surface states (Figure S4). The d*I*/d*V* maps at the energy near $E_F$ (e.g., 0 meV and 400 meV, Figure 1(e)) show strong intensity contrast between two surfaces, further demonstrating the distinct electronic characteristics for the Sn and Au surfaces.

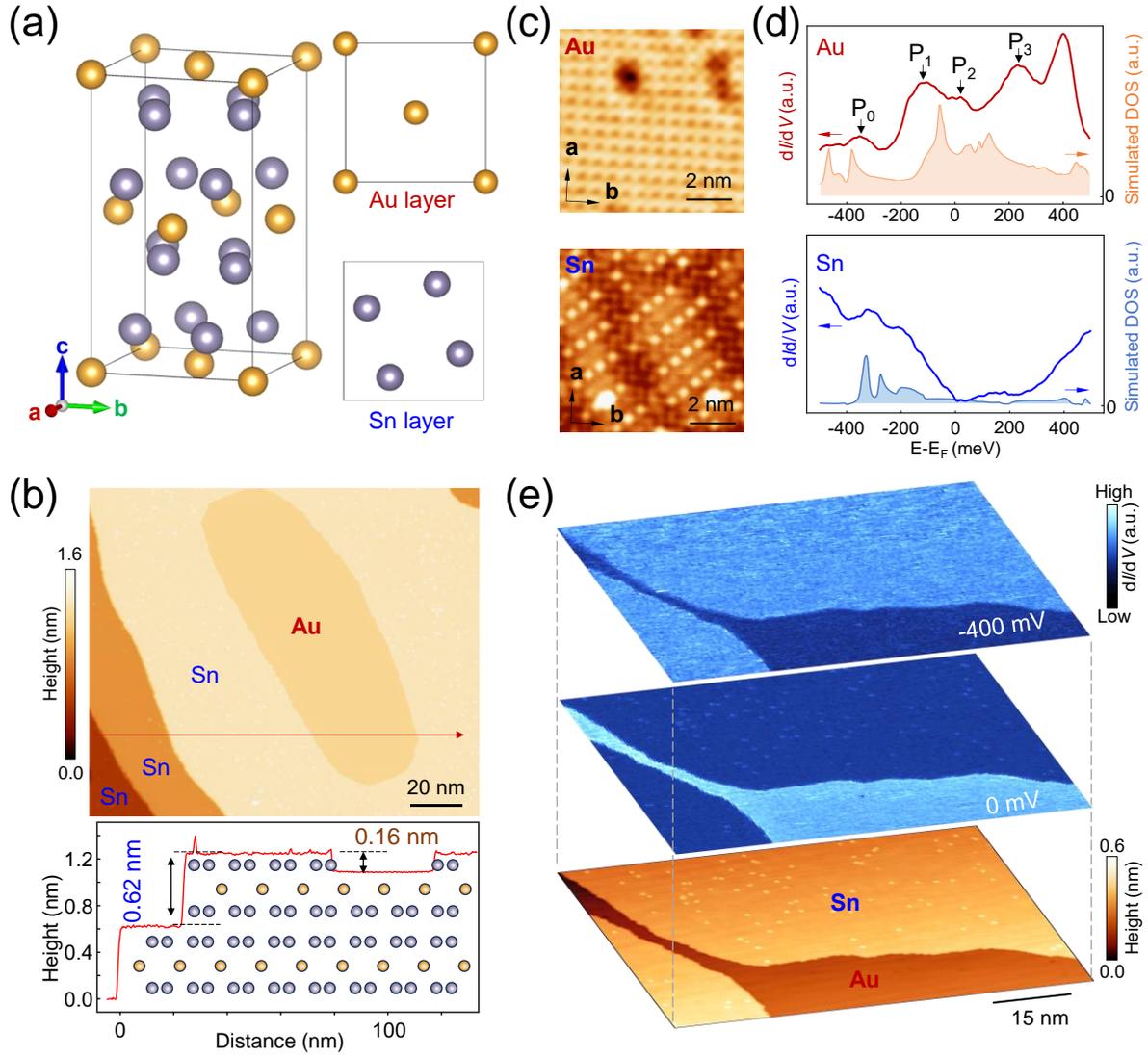

**Figure 1.** Atomic model, structural and electronic properties of Au and Sn surfaces of AuSn$_4$. (a) Schematic structure of AuSn$_4$ crystal with Au atoms in yellow, Sn atoms in gray. The solid lines represent a unit cell. (b) Upper panel: Large-scale STM image of the as-cleaved AuSn$_4$, showing the Sn and Au terraces, respectively ($V_s$ = -90 mV, $I_t$ = 1 nA). Lower panel: Line profile across the red arrow in the STM image, showing that the height between the two adjacent An and Sn terraces is ~0.16 nm and the height between two neighboring Sn terraces is ~0.62 nm. The schematic of side view of AuSn$_4$ atomic model is overlaid for clarifying the Au and Sn terminated surfaces. (c) Atomically resolved STM images, showing the square lattice of the An-terminated (Upper, $V_s$ = -100 mV, $I_t$ = 1 nA) and the unidirectional CDW on the Sn-terminated surfaces (Lower, $V_s$ = -50 mV, $I_t$ = 1 nA), respectively. (d) Spatially averaged d$I$/d$V$ spectra obtained at 4 K and calculated density of states results for two surfaces, showing distinct electronic states. $V_s$ = -500 mV, $I_t$ = 1 nA, $V_{mod}$ = 10 mV. (e) STM topography and corresponding d$I$/d$V$ maps $g$(-400 mV,$r$) and $g$(0 mV,$r$), revealing the significant differences in the surface states of Au and Sn. $V_s$ = -800 mV, $I_t$ = 2 nA, $V_{mod}$ = 10 mV.

To further investigate the origin of distinct surface states for the two surfaces, we apply quasiparticle interference (QPI) imaging, a powerful tool for probing the dispersion relations of the surface bands[43]. The Sn surface, decorated by moderate concentration of adatoms, exhibits much weaker QPI patterns (Figure S5), consistent with the metallic bulk features observed in the d$I$/d$V$ spectrum. In contrast, the d$I$/d$V$ maps on the Au surface reveal strong QPI patterns around vacancies (Figure 2(a)). The Fast Fourier transform (FFT) of representative d$I$/d$V$ map taken at low energy of -300 meV exhibits a ring shape (blue dashed circle in Figure. 2(b)). With increasing energy, the ring-shape QPI patterns increase in size. Around 50 meV, an additional small circular pattern (red arrow in Figure 2(b)) appears around $q$=0. Different from larger QPI ring where the scattering vector only exist at the edge of ring, the new QPI circle exist scattering vectors inside the whole circle, which may be due to multiple scattering vectors from various bands. At higher energy of 200 meV, the circle QPI pattern enlarges into a larger ring and merges with the outer ring pattern, forming a donut-like QPI pattern (green arrow in Figure 2(b)).

The full energy evolution of QPI patterns observed on the Au surface is captured in the energy-momentum cut along one lattice direction, i.e. Γ-X (Figure 2(c)). The most prominent dispersing parabolic curve (dashed blue curve in Figure 2(c)), which grows outwards from $q$=0 with increasing energy, corresponds to the expending ring-like QPI patterns. At the energy range between 50 and 300 meV, new dispersion features appear at lower $q$ inside the parabolic dispersion, which corresponds to the growing circular and donut-like QPI patterns. In general, the QPI patterns consist of a complicated mixture of scattering processes that involve both trivial and nontrivial surface bands.

The identification of the respective surface bands that generate the QPI patterns is achieved by detailed comparison with calculations of surface bands. In the calculated surface band, the Fermi energy (zero sample bias in STM) is set consistently for both surface terminations to lie 150 meV above the neutrality point, based on fitting the respective QPI evolution with calculations. The comparisons between QPI evolution and calculated surface band of Au terminated surface along Γ-X reveal that the prominent dispersing parabolic band is the conventional Rashba splitting surface band (Figure 2(d)), similar to the Shockley surface states on the (111) surface of Au single crystal[44]. The band minimum of Rashba splitting surface band between -300 and -400 meV is comparable to the peak $P_0$ in d$I$/d$V$ spectrum (Figure 1(d)).

Significantly, the calculated Au terminated surface band demonstrates that the coupling between an electron-like Rashba band and a hole-like Rashba band leads to an unconventional Rashba Fermi surface, where two circles from two different bands possess the same spin chirality arises[29] (Figure 2(e)). The scattering within such unconventional Rashba Fermi surface gives rise to the emergent circular QPI patterns which grow into a donut-like pattern. The unconventional Rashba band contribute to the peak near $E_F$ in the density of state, which corresponds to the observed $P_2$ in the d$I$/d$V$ spectrum (Figure 1(d)).

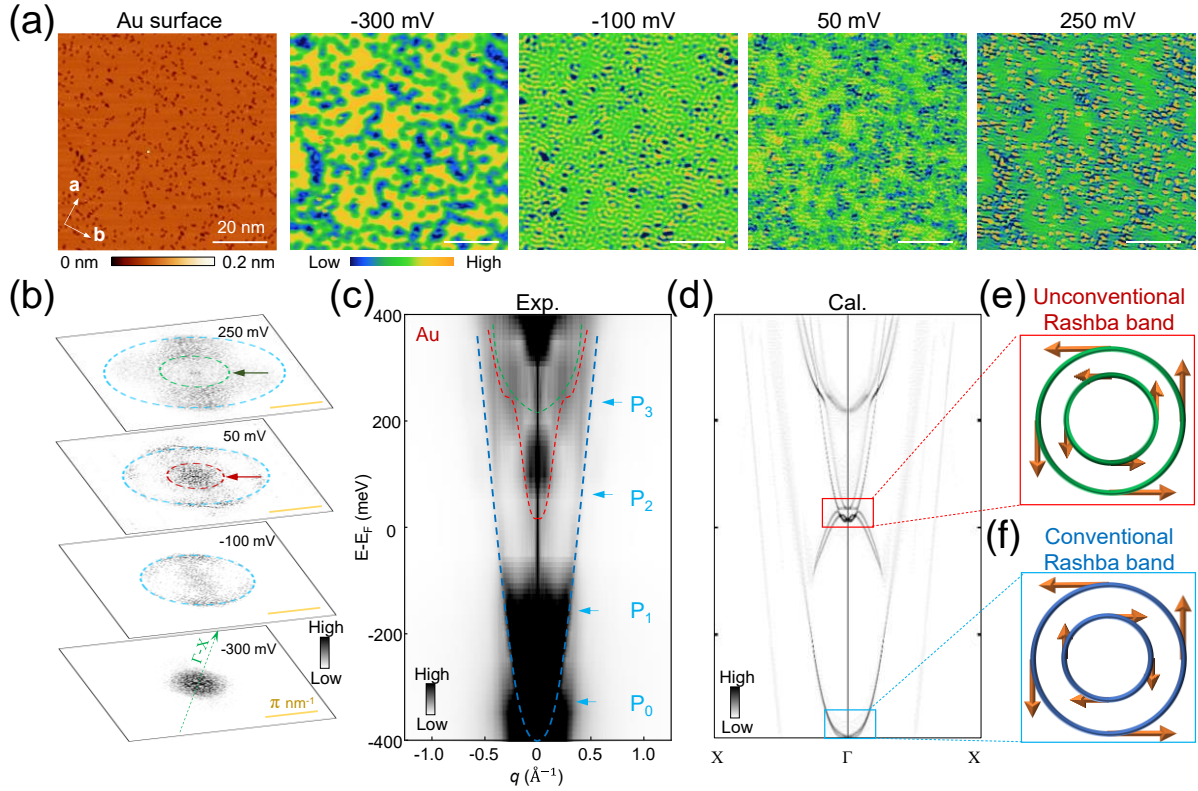

**Figure 2.** The QPI patterns and calculated electronic states of Au terminated surfaces of $AuSn_4$. (a) Topography and a series of d$I$/d$V$ maps obtained on the Au-terminated surfaces, showing the standing wave-like patterns around the impurities. Topography: $V_s$ = -400 mV, $I_t$ = 1 nA, maps: $I_t$ = 5 nA, $V_{mod}$ = 10 mV, $V_s$ is shown on each panel. (b) Fast Fourier transforms of a series of d$I$/d$V$ maps at different energies shown in (a), displaying the energy dependent QPI patterns. (c) Experimental FT cut of $g(E, \boldsymbol{q})$ along Γ-X, showing the surface state dispersions. (d) Simulated cut of $g(E, \boldsymbol{q})$ along Γ-X direction. (e-f) Spin textures of two Fermi circles from the Rashba bands in (d), showing conventional and unconventional Rashba bands, respectively.

The termination-dependent states near $E_F$ naturally raise the question of whether there are any termination-dependent superconductivity in $AuSn_4$. To explore this, we investigate the low-energy density of state at an electron temperature of 0.17 K, significantly lower than the superconducting critical temperature ($T_c$). We observe a superconducting gap symmetric to the $E_F$ on both An and Sn surfaces (Figure 3(a)). By subtracting the normal electronic surface states, we find that two surfaces exhibit almost the same Bardeen–Cooper–Schrieffer like superconducting gap of $\Delta \sim 0.43$ meV (Figure 3(b)). Moreover, the superconducting gap is not affected by adatoms or vacancies (Figure S6), suggesting $s$-wave superconductivity. The only notable difference is that the Sn surface possesses a more prominent coherence peak (Figure S7). The temperature dependent d$I$/d$V$ spectra (Figure 3(c), (d)) show the $T_c \sim 2.21$ K for both surfaces, which agrees well with transport measurements[37-40]. Magnetic field dependent d$I$/d$V$ curves (Figure 3(e-h)) show the critical field $H_{c2}$ of ~ 0.15 T for in-plane field ($H//ab$) and ~0.04 T for out-of-plane ($H//c$).

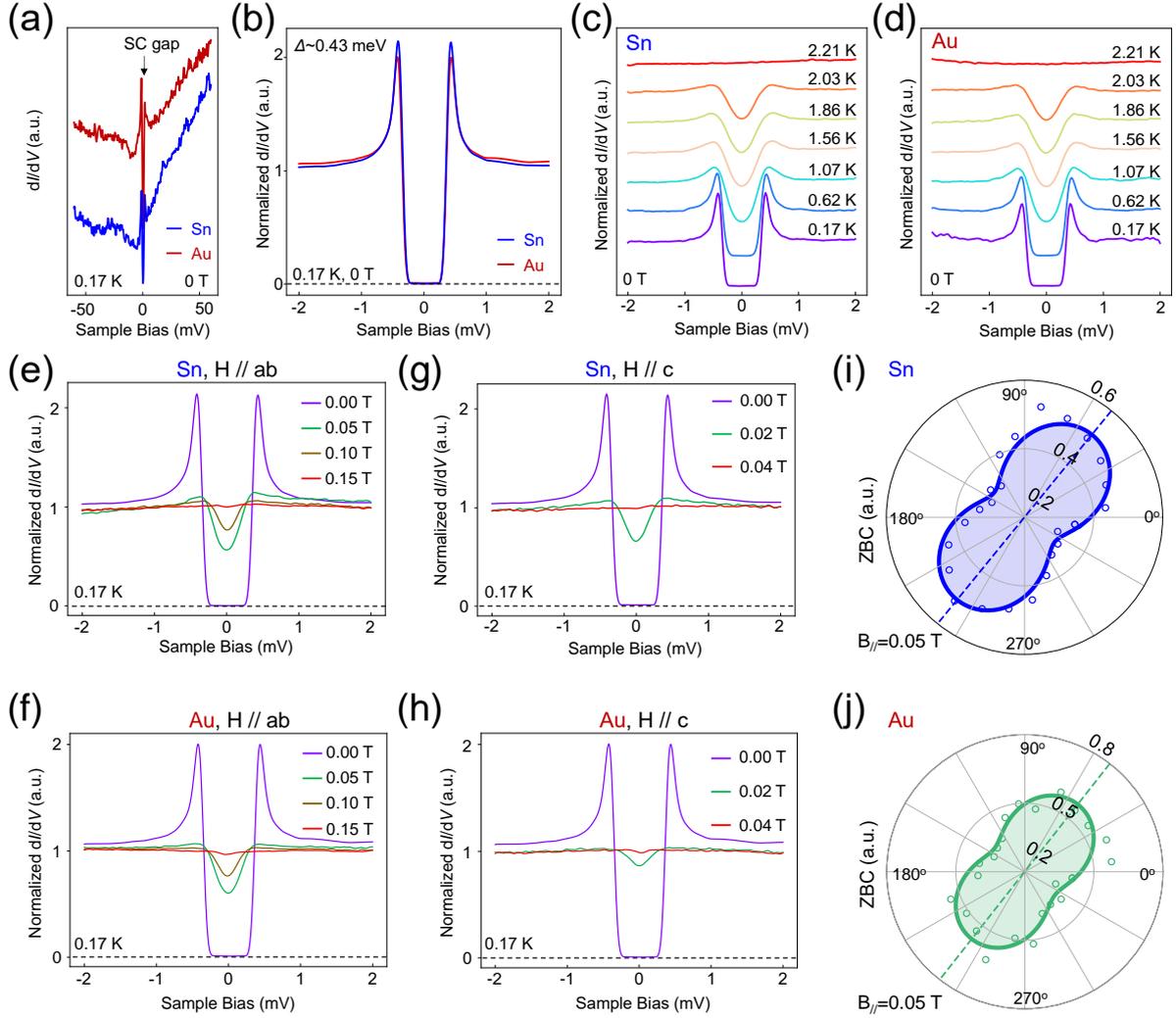

**Figure 3. Temperature and field evolution of superconducting gap on Au and Sn terminated surfaces.** (a) The spatially averaged d$I$/d$V$ spectra obtained at an electron temperature of 0.17 K, showing an emergent superconducting gap symmetric to the E$_F$ on both surfaces ($V_s$ = -90 mV, $I_t$ = 1 nA, $V_{mod}$ = 1 mV). (b) Low-energy d$I$/d$V$ spectra, showing a superconducting gap $\Delta$~0.43 meV with different coherence peak strength on Au and Sn surfaces ($V_s$ = -2 mV, $I_t$ = 1 nA, $V_{mod}$ = 50 μV). (c-h) Temperature and magnetic field dependence of d$I$/d$V$ spectra on Au surface ($V_s$ = -2 mV, $I_t$ = 1 nA, $V_{mod}$ = 50 μV), showing a similar critical temperature $T_c$ ~2.21 K and critical magnetic field $H_{c2}^{ab}$ ~ 0.15 T, $H_{c2}^{c}$ ~ 0.04 T on both surfaces. The temperature used here is electronic effective temperature. (i-j) Polar plot of zero-bias conductance of d$I$/d$V$ spectra measured under in-plane magnetic field at different $\theta$ on the Sn and Au surface respectively, showing the obvious two-fold symmetry. Note that we measure the d$I$/d$V$ spectra at locations far away from the vortex under in-plane magnetic field. The spectra in (c) and (d) are vertically shifted for clarity.

Although the superconducting gap and corresponding critical parameters are identical for both surfaces, the gap depth exhibit significant differences under the same magnetic field (Figure S8). We measure the in-plane field dependent d$I$/d$V$ spectra along a field direction perpendicular to that in Figure 3(f) and obtain the critical field of 0.10 T (smaller than that of 0.15 T in Figure 3(f)), indicating there may exist an angular dependence for the superconducting gap (Figure S8). To investigate the intriguing symmetry-breaking superconducting states, we collect d$I$/d$V$ spectra at an electron temperature of 0.17 K by applying an in-plane magnetic field of B$_{//}$=0.05 T and study the superconducting gap as a function of $\theta$ defined as the azimuthal angle of B$_{//}$ with respect to the $a$-axis of crystalline lattice. To avoid the effect of field induced vortices, we collect the d$I$/d$V$ spectra at locations away from the Abrikosov vortex cores (Figure S9 and Figure 3(i), (j)). Although the differences in the peak-to-peak distance for the superconducting gap are difficult to be identified, zero-bias conductance (ZBC) inside the superconducting gap shows significant difference with rotating the angles. The angular dependence of ZBC based on an extensive dataset exhibits twofold symmetry, consistent with the symmetry-breaking superconducting states in the previous transport experiments[40].

Given that angle-dependent superconductivity in STM/S measurements show similar C$_2$ features with transport experiments, the next question is whether previously proposed surface $p$-wave superconductivity and MZMs[40] exist. To this end, we apply external magnetic field perpendicular to the surfaces ($B_z$) and map the field induced Abrikosov vortices in a large-area surface region containing both Au and Sn surfaces (Figure 4(a), details see Experimental Method). At $B_z$= 0.02 T, hexagonal vortex lattice with lattice constant of ~350 nm is observed (Figure 4(a), (b)). However, the vortex cores show higher ZBC intensity at Au surface than the Sn surface. Such intriguing phenomenon is more clearly illustrated in the d$I$/d$V$ linecuts across vortices. A sharp zero bias conductance peak (ZBCP) with long decay distance emerges at the vortex core on the Au surface, while a broad weak ZBCP with short decay distance appears on the Sn surface (Figure 4(c), (d)). The distinct ZBCP spatial evolution between Au and Sn surfaces is robust against surface regions (Figure S10) and magnetic field (Figure 4(e), (f) and Figure S11) by using different tip states, supporting that the distinction originates from intrinsic quantum states at Au terminated surfaces rather than the extrinsic tip or impurity effects. The isotropy of the vortex is also confirmed by showing d$I$/d$V$ spectra linecuts along another vertical path (Figure S12). The termination dependent vortex bound states are even observed within a single vortex across a narrow Au terrace sandwiched by two Sn terraces where the Au terrace show much higher density of states than the one of Sn terrace (Figure 4(g), (h)).

For the case of the vortex core on a conventional superconductor, Caroli-de Gennes-Matricon bound states (CBSs) emerge at discrete energy levels $E_\mu = \mu \Delta^2/E_F$ ($\mu$ is half-integer angular momentum number and equals ±1/2, ±3/2 …)[45], while MZM appears when topological surface states combined with superconductivity[46]. Quasiparticles gain an additional half-integer angular momentum inside a vortex and $\mu$ becomes integer quantized[47]. For most superconductors with relative low $T_c$, the energy spacing between discrete energy levels is within the resolution of most experimental equipment and cannot be discerned, which terms as not reaching quantum condition. Therefore, researchers refer an "X"-type splitting along the path across a vortex core as the conventional CBSs case[48] and a "Y"-type splitting representing the MZM extending in real space at a finite length as the MZM case[12, 49]. Here for AuSn$_4$, vortex on the Sn surface show a trivial "X"-type like feature and vortex on the Au surface show a non-trivial "Y"-type like feature with the prominent ZBCP extending long spatial distance. Moreover, the ZBCP intensity measured at the vortex core of Au surface decreases with increasing temperature, and becomes extremely weak at 0.62 K and total invisible at 0.76 K, as shown in Figure 4(i). A CBS peak will exhibit simple Fermi-Dirac broadening and persist to higher temperature up to $T_c/2$ (about 1.1 K). Our observation contradicts this expectation and indicates that the fast suppression is possibly owing to the poisoning of MZMs by thermally excited quasiparticles.

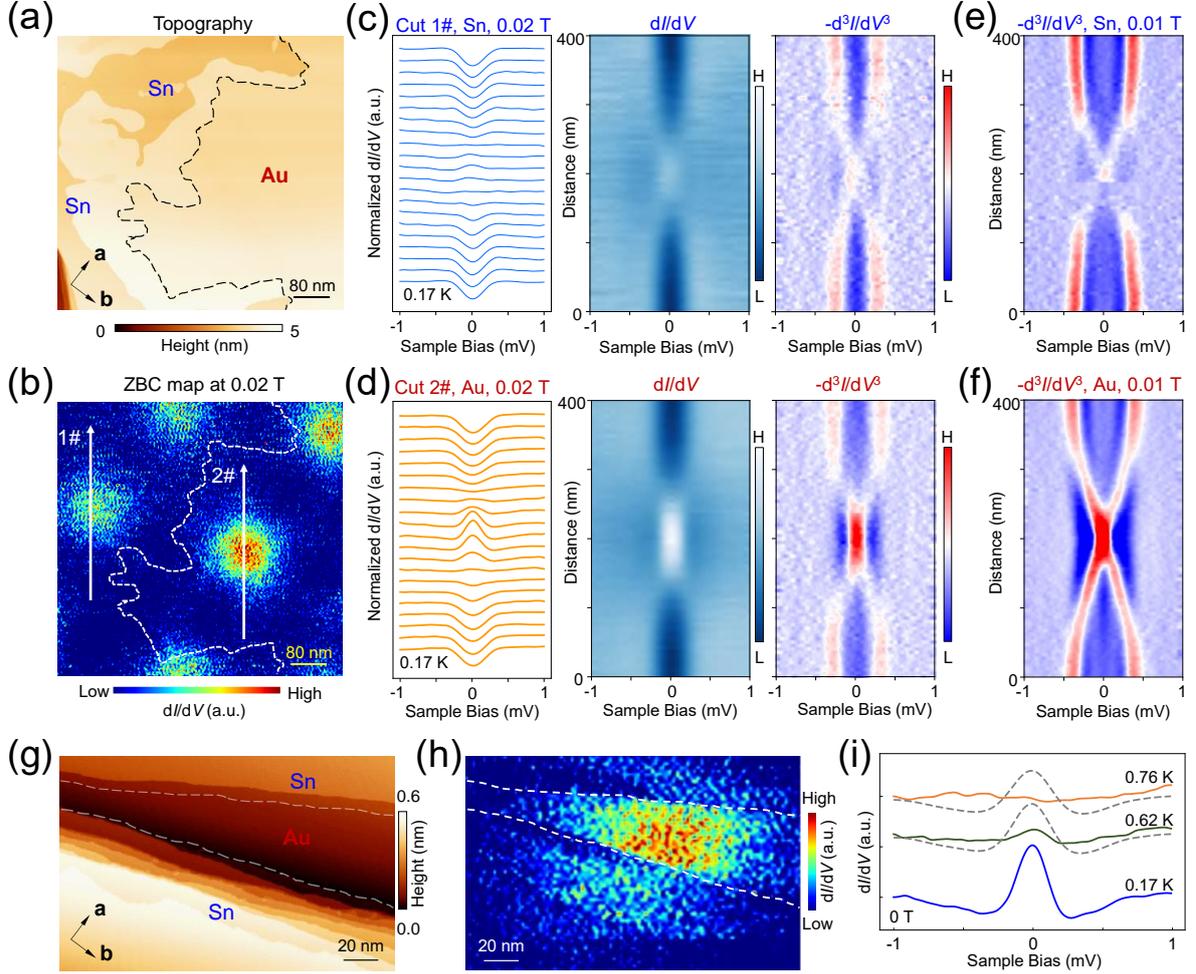

**Figure 4. Distinct vortex bound states on Au and Sn surfaces.** (a) STM image of a large terrace area ($V_s$ = -90 mV, $I_t$ = 100 pA). (b) Zero bias conductance (ZBC) map of (a) at $B_z$ = 0.02 T, showing vortex lattice and distinct ZBC core intensity on Sn-terminated and Au-terminated surface ($V_s$ = -2 mV, $I_t$ = 500 pA, $V_{mod}$ = 50 μV). (c-d) The d$I$/d$V$ linecuts and corresponding negative second differential curves along the two white solid line with arrows in (b), showing different vortex core states at Au and Sn surfaces. (e-f) Negative second differential curves of d$I$/d$V$ linecuts along vortex in the same region at $B_z$ = 0.01 T. Details of the vortex location and linecuts is shown in Fig. S11. (g) STM image of a terrace area ($V_s$ = -90 mV, $I_t$ = 100 pA). (h) ZBC map of (g) at $B_z$ = 0.03 T, showing a vortex appearing at a narrow Au surface sandwiched by two Sn regions. (i) Temperature evolution of ZBCPs in a vortex core on Au surface. The grey curves are numerically broadened 0.17 K data at each temperature. The temperature used here is electronic effective temperature.

The density of states at Sn terminated surface mainly contribute from bulk AuSn$_4$ while the Au terminated surface show strong unconventional surface state near E$_F$. In addition, the coherence length subtracted from the vortex bound states of Sn surface (~102 nm) is comparable to the one of bulk superconductivity (~97 nm), but shorter than the coherence length of Au surface (~75 nm) (Figure S12 (e-f), details see Experimental Method). Therefore, we attribute the signature of MZM on the Au terminated surface to the existence of unconventional Rashba Fermi surface at the superconducting state.

The surface states on the Au surface are much stronger than those on the Sn surface and include unconventional Rashba splits. At the crystal surface, due to the breaking of inversion symmetry, a typical Rashba band arises and meanwhile the mixing of spin singlet and triplet superconductivity is induced. For the conventional two-band Rashba system, its Fermi surface is composed of two circles which have opposite chirality of spin texture when the two circles belong to different bands. The superconducting order parameter projected to two Fermi circles has amplitude $\Delta_+$ and $\Delta_-$ for the inner and outer circles, respectively. It has been studied that if $\Delta_+$ and $\Delta_-$ have the same sign, the singlet component is dominated and accordingly the system is topological trivial; On the contrary, if $\Delta_+$ and $\Delta_-$ have different signs, the triplet component is dominated and the system is topologically nontrivial[50]. In real, triplet-dominated state is hard to get and one of the possible solutions is to delete one of the spin split bands by external magnetic field or by the exchange splitting due to the proximity to a ferromagnet[51]. However, when the orbital degree of freedom is considered, things become more interesting. As for our AuSn$_4$ system, the symmetry of its surface can be described by point group $C_{4v}$ and split by crystal field and spin-orbit coupling, both s and $p_z$ orbital electrons can form a state belonging to $\Gamma_6$ irreducible representation. Each $\Gamma_6$ state can solely constitute a simple Rashba band, whereas when these two Rashba bands close to each other in momentum space, the interorbital hybridization is allowed by symmetry. The consequence is that a band kink is generated and unconventional Rashba Fermi surface where two circles from two different bands possess the same spin chirality arises[29]. Accordingly, both inter- and intraorbital pairing can emerge. For AuSn$_4$, six possible pairing ways form three order parameters[40]. The stronger interorbital interaction drives system to enter interorbital triplet superconductivity[52]. In this case, it is easy to find that the superconducting amplitude on two Fermi circles has different signs, which is topological nontrivial. Thus, the robust ZBCP on the Au surface may serve as the evidence of the presence of MZM. Further higher energy resolution studies are needed to fully identify the MZMs on Au surfaces.

In summary, we identify two types of terminated surfaces in high-quality AuSn$_4$ single crystal, exhibiting totally different electronic surface states and vortex-core states. The Au surface possesses stronger surface states including unconventional Rashba bands and hosts vortices with robust zero energy bound states, which may serve as evidence of MZMs. Our findings inspire the synthesis of superconductors consisting of noble metal elements, such as gold, can achieve terminated surfaces with strong surface states based on proper cleavage conditions, which is analogous to the induced superconductivity on Au thin film[53, 54]. These types of superconductors provide a potential platform for distinguishing of MZM from trivial vortex bound states and braiding manipulation of MZM.

# EXPERIMENTAL METHOD

**Single-Crystal Growth of AuSn$_4$.** The single crystals of AuSn$_4$ were synthesized by the self-flux method with excess Sn[55, 56]. The sources are high-purity Au wire and Sn pellets which weighed with a molar ratio of 12:88 in the glove box. Then mixture of the sources was transferred to an alumina crucible and sealed in an evacuated quartz tube. The quartz tube was heated to 850 °C by a box furnace and held for 24 hours, followed by cooling to 310 °C for 11 h and then slowly cooled down to 230 °C with a rate of 0.5 °C/h. Subsequently, the extra Sn flux is removed by centrifuge at 230 °C. Large AuSn4 single crystals with a silvery luster were obtained. Structure characterization and transport measurements demonstrate the high quality crystalline of as-grown samples (Figure S1).

**Sample Characterization of the AuSn$_4$.** The XRD pattern of as-grown AuSn$_4$ single crystals was collected using a Rigaku SmartLab SE X-ray diffractometer with Cu K$_\alpha$ radiation ($\lambda$ = 0.15418 nm) at room temperature. Energy-dispersive X-ray spectroscopy (EDX) were performed using a HITACHI S5000 with an energy dispersive analysis system Bruker XFlash 6|60. The transport measurements were realized by the physical property measurement system (PPMS, Quantum Design Inc.) with 16 T superconducting coils and a temperature range of 400 K to 1.8 K.

**Sample cleavage.** The samples are cleaved in situ at room temperature and immediately transferred to an STM chamber. We noted that Au terminated surface are rarely reported in previous work. The cleavage temperature may contribute to the formation of two different surfaces. We cleave four samples of AuSn$_4$, and find two kinds of surfaces on two samples cleaved at room temperature but only the Sn surface on the other samples cleaved at liquid nitrogen temperature (~77K). Higher temperature may offer the relaxation energy for the covalent bonding within the tri-layer to break up and the Au layer becomes the terminated surface.

**Scanning Tunneling Microscopy/Spectroscopy.** The samples used in the experiments were cleaved in situ at room temperature and immediately transferred to an STM chamber. Experiments were performed in an ultrahigh vacuum (1×10$^{-10}$ mbar) ultra-low temperature STM system ($T_{\text{base}}$~0.03 K, $T_{\text{electron}}$~0.17 K) equipped with 9-2-2 T magnetic field. All the scanning parameters (setpoint voltage and current) of the STM topographic images are listed in the captions of the figures. Unless otherwise noted, the differential conductance (d$I$/d$V$) spectra were acquired by a standard lock-in amplifier at a modulation frequency of 973.1 Hz. Tungsten tip used in the experiment was fabricated via electrochemical etching and calibrated on clean Au (111) surface prepared by repeated cycles of sputtering with argon ions and annealing at 500 °C.

**Coherence Length analysis.** We estimate the coherence length ξ value equals 97 nm by using the formula $H_{c2} = \Phi_0/2\pi\xi_{GL}^2$ ($\Phi_0$ denotes the magnetic flux quanta) based on Ginzburg-Landau theory and H$_{c2}$≈345 Oe from the transport measurement. Therefore, we select a large field of view (about 1 μm$^2$) consisting of both Au and Sn surfaces (Figure 4(a)).

The distinct vortex bound state behavior strongly correlated to the superconducting nature between the two terminated surfaces. Thus we extract the zero-bias conductance $Z$ as a function of distance from the vortex core $r$ and fit the $Z(r)$ by using decay function $Z(r) = A + Bexp(-\frac{r}{\xi})$, where A and B is fitting parameters[57, 58]. From the fitting we can get ξ values equal 101.84 ± 8.50 nm and 74.87 ± 3.70 nm for vortex on Sn and Au surfaces, respectively. We find that the ξ value on the Sn surface is closer to the value we obtain from the Ginzburg-Landau theory, which may indicate the vortex on Sn reflects the bulk nature while the vortex on the Au surface is affected by the surface states and reflects some two-dimensional surface nature.

**DFT Calculations.** First-principles calculations were performed by density functional theory (DFT) using the Vienna ab initio simulation package (VASP)[59, 60]. The plane-wave basis with an energy cutoff of 350 eV was adopted. The electron-ion interactions were modeled by the projector augmented wave potential (PAW)[61] and the exchange-correlation functional was approximated by the Perdew-Burke-Ernzerhof-type (PBE) generalized gradient approximation (GGA)[62]. The structural relaxation for optimized lattice constants and atomic positions was performed with an energy (force) criterion of $10^{-8}$ eV (0.01 eV/Å) and by using the DFT-D3[63] method to include van der Waals corrections. Surface state calculations were performed with the WannierTools package[64], based on the tight-binding Hamiltonians constructed from maximally localized Wannier functions (MLWF)[65].